\documentclass[conference]{IEEEtran}
\IEEEoverridecommandlockouts
\usepackage{cite}
\usepackage{amsmath,amssymb,amsfonts}
\usepackage[ruled,noend,linesnumbered]{algorithm2e} 

\usepackage{algpseudocode}
\usepackage{graphicx}
\usepackage{xcolor}
\usepackage{graphicx}
\usepackage{textcomp}
\usepackage{xcolor}
\usepackage{enumitem}
\usepackage{url}
\usepackage{bm}
\usepackage{makecell}

\definecolor{red}{rgb}{0,0,0} 
\definecolor{blue}{rgb}{0,0,1} 
\definecolor{dispatchers1}{rgb}{0.412,0.392,0.482} 
\definecolor{orchestrator1}{rgb}{0.69,0.478,0.596} 
\definecolor{dispatchers2}{rgb}{0.380,0.227,0.545} 
\definecolor{orchestrator2}{rgb}{0.525,0.220,0.271} 

\def\BibTeX{{\rm B\kern-.05em{\sc i\kern-.025em b}\kern-.08em
    T\kern-.1667em\lower.7ex\hbox{E}\kern-.125emX}}

\begin{document}
\title{\textit{EdgeMatrix}: A Resources Redefined Edge-Cloud \\ System for Prioritized Services}
\author{
    \IEEEauthorblockN{Yuanming Ren\IEEEauthorrefmark{2}, Shihao Shen\IEEEauthorrefmark{2}, Yanli Ju\IEEEauthorrefmark{2}, Xiaofei Wang\IEEEauthorrefmark{2}, Wenyu Wang\IEEEauthorrefmark{5},
		Victor C.M. Leung\IEEEauthorrefmark{3}\IEEEauthorrefmark{4}}
    \IEEEauthorblockA{\IEEEauthorrefmark{2}TANKLab, College of Intelligence and Computing, Tianjin University, Tianjin, China}
    \IEEEauthorblockA{\IEEEauthorrefmark{5}Shanghai Zhuichu Networking Technologies Co., Ltd., China}
    \IEEEauthorblockA{\IEEEauthorrefmark{3}College of Computer Science and Software Engineering, Shenzhen University, Shenzhen, China}
    \IEEEauthorblockA{\IEEEauthorrefmark{4}Dept. of Electrical and Computer Engineering, the University of British Columbia, Vancouver, Canada}
	\IEEEauthorblockA{\{renyuanming, shenshihao, yanliju,  xiaofeiwang\}@tju.edu.cn, wayne@pplabs.org, vleung@ieee.org}
	\vspace{-1.5em}
	\thanks{This work was supported in part by the National Key Research and Development Program of China under Grant No. 2019YFB2101901; the National Science Foundation of China under Grant No. 62072332; Corresponding author: \textit{Xiaofei Wang}.}
}

\maketitle

\begin{abstract}

The edge-cloud system has the potential to combine the advantages of heterogeneous devices and truly realize ubiquitous computing. However, for service providers to {\color{red}guarantee the Service-Level-Agreement (SLA) priorities}, the complex networked environment brings inherent challenges such as multi-resource heterogeneity, resource competition, and networked system dynamics. In this paper, we design a framework for the edge-cloud system, namely \textit{EdgeMatrix}, to maximize the throughput while {\color{red}guaranteeing various SLA priorities}. First, \textit{EdgeMatrix} introduces Networked Multi-agent Actor-Critic (NMAC) algorithm to redefine physical resources as logically isolated resource combinations, i.e., resource cells. Then, we use a clustering algorithm to group the cells with similar characteristics into various sets, i.e., resource channels, for different channels can offer different SLA guarantees. Besides, we design a multi-task mechanism to solve the problem of joint service orchestration and request dispatch (JSORD) among edge-cloud clusters, significantly reducing the {\color{red}runtime} than traditional methods. To ensure stability, \textit{EdgeMatrix} adopts a two-time-scale framework, i.e., coordinating resources and services at the large time scale and dispatching requests at the small time scale. The real trace-based experimental results verify that \textit{EdgeMatrix} can improve system throughput in complex networked environments, reduce SLA violations, and significantly reduce the {\color{red}runtime} than traditional methods.

\end{abstract}

\section{Introduction}
\label{sec:Introduction}

\subsection{Background and Problem Statement}
\label{subsec:Background and Problem Statement}

With the explosion of networked devices, centralized mobile network architecture is facing many challenges. According to the GSMA's \textit{The Mobile Economy 2020} report, IoT connections will reach almost 25 billion globally by 2025, up from 12 billion in 2019\cite{GSMA2020}. As a result, the traditional cloud computing paradigm is hard to cope with:(\textit{$\romannumeral1$}) a large number of computing tasks generated by the massive networked devices are delivered to the cloud center, which brings a severe challenge to cloud computing capability; (\textit{$\romannumeral2$}) the long transmission distance between the networked devices and the cloud center, which is difficult to meet the requirements of low latency services, e.g.,  automated driving.

To address the above issues, the emergence of edge computing\cite{shi2016} has the potential to guide the development of next-generation network architecture. Compared with cloud computing, the advantages of edge computing are mainly shown in two aspects: (\textit{$\romannumeral1$}) widely distributed edge computing nodes can handle a large number of computing tasks, relieving the pressure on the backbone network; (\textit{$\romannumeral2$}) most services are processed near the edge, reducing the data transmission delay, and only services that cannot be processed at the edge will be uploaded to the cloud center.

Therefore, edge computing can effectively solve many problems such as low real-time and work inefficiency in traditional cloud computing.  Simultaneously, cloud computing can provide great computing power and massive storage for distributed edge computing nodes.
The complementary characteristics of edge computing and cloud computing accelerate the deep collaboration between each other and gradually evolve into edge-cloud system\cite{wang2020}. Unlike the one-fits-all cloud computing paradigm, the widely distributed edge nodes and mutually heterogeneous edge clusters in the edge-cloud system bring significant opportunities and also challenges to provide users reliable service\cite{tran2018}.

\begin{figure}[t]
	\centering
	\includegraphics[width=8.85 cm]{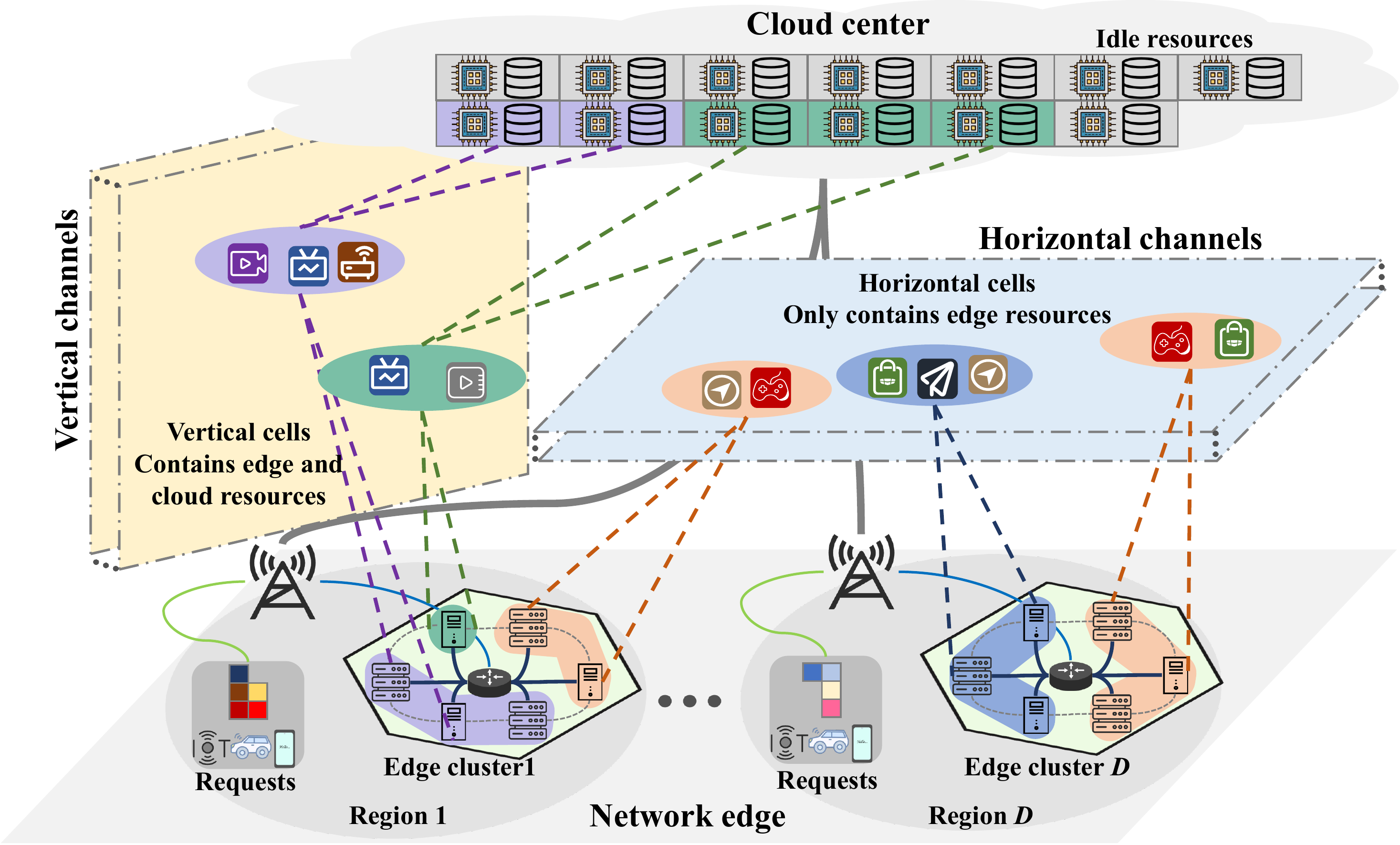}
	\setlength{\abovecaptionskip}{-0.5cm} 
	\caption{Architechture of \textit{EdgeMatrix}.}
	\label{fig:SystemArchitechture}
	\vspace{-1.6em}
\end{figure}

\subsection{\color{red}Motivation and Challenges }
\label{subsec:Motivation and Challenges }

In the cloud computing paradigm, service providers can provide reliable services to users based on Service-Level-Agreement (SLA)\cite{wu2011}.
In this paper, we {\color{red} propose \textit{EdgeMatrix} that can} provide strong SLA assurance for various user services in the complex networked environment of the edge-cloud system based on the idea of SLA in cloud computing.

Although providing reliable service for users based on SLAs in edge-cloud systems can significantly improve system efficiency, three inherent challenges still need to be faced in the specific implementation process.
(\textit{$\romannumeral1$}) \textit{\textbf{Multi-resource heterogeneity}}: Geographically distributed edge nodes have different computing capabilities, communication capabilities and system architectures;
(\textit{$\romannumeral2$}) \textit{\textbf{Resource competition}}: Different types of services have various resource requirements, which causes resource competition among different services and thus affects the service efficiency of requests;
(\textit{$\romannumeral3$}) \textit{\textbf{Networked system dynamics}}: Due to the random fluctuations in user demand and networked devices, the request load and available resources of the networked system are in a constant dynamic change.
Therefore, there is an urgent need for the resource-redefined architecture to satisfy the SLAs of users under the inherent challenges orient to edge-cloud systems.


\vspace{-0.2em}
\subsection{Technical Challenges and Solutions}
\label{subsec:Technical Challenges and Solutions}
\vspace{-0.3em}


In this paper, to better cope with the three inherent challenges in the edge-cloud system, our work focuses on resource customization, service orchestration, and request dispatch.

\textit{\textbf{Resource customization}}. The multi-resource heterogeneity of the networked system poses a severe problem for providing users reliable service in edge-cloud systems {\color{red}because heterogeneous edge nodes increase the uncertainty of service orchestration and request dispatch.} It is challenging to design traditional methods to consider the massive heterogeneous nodes in the system, i.e., a large number of constraints may cause algorithms complicated and even unsolvable\cite{papadimitriou1998}. Therefore, we introduce multi-agent deep reinforcement learning (MADRL) algorithms to provide customized isolated resources for various user services in the edge-cloud system. \textbf{Specially, we customize the resources of edge-edge nodes (Horizontal) and edge-cloud nodes (Vertical) to form logically isolated resource combinations called \textit{resource cells} in edge-cloud systems.} Biologically, cells at different locations have different functions, and each cell has an isolated space. \textbf{We further call the set of cells with similar characteristics (resources, latency, etc.) a \textit{resource channel}, which means that each resource channel has its corresponding {\color{red}SLA priority}.} Macroscopically, resource channels can also be divided into two categories, i.e., horizontal or vertical. As shown in Fig. \ref{fig:SystemArchitechture}, we call this framework \textit{EdgeMatrix}.

\textit{\textbf{Service orchestration}}. Resource competition among services can lead to a decrease in the number of requests that are successfully served in the system without violating SLA priority, i.e., a reduction in throughput\cite{zhang2019}. Imagine a scenario where one service takes up most of the memory resources on a node, under which orchestration of other services is severely adversely affected, even if they require merely a few memory resources. Therefore, we should orchestrate the services reasonably by \textit{EdgeMatrix} to reduce the negative impact of resource competition. 

\textit{\textbf{Request dispatch}}. The dynamic of networked systems poses a significant challenge for the adaptability of dispatch algorithms\cite{hu2019}. Request dispatch is the last link to determine whether requests can be successfully served. The design of the request dispatch algorithm plays a crucial role in system robustness in the face of networked system dynamics. Specifically, to ensure the system's stability, we adopt a two-time-scale framework, which performs resource customization and service orchestration sequentially at the large time scale (frame) and requests dispatch at the small time scale (slot).

\subsection{Main Contributions}
\begin{figure}[t]
	\centering
	\includegraphics[width=8.85 cm]{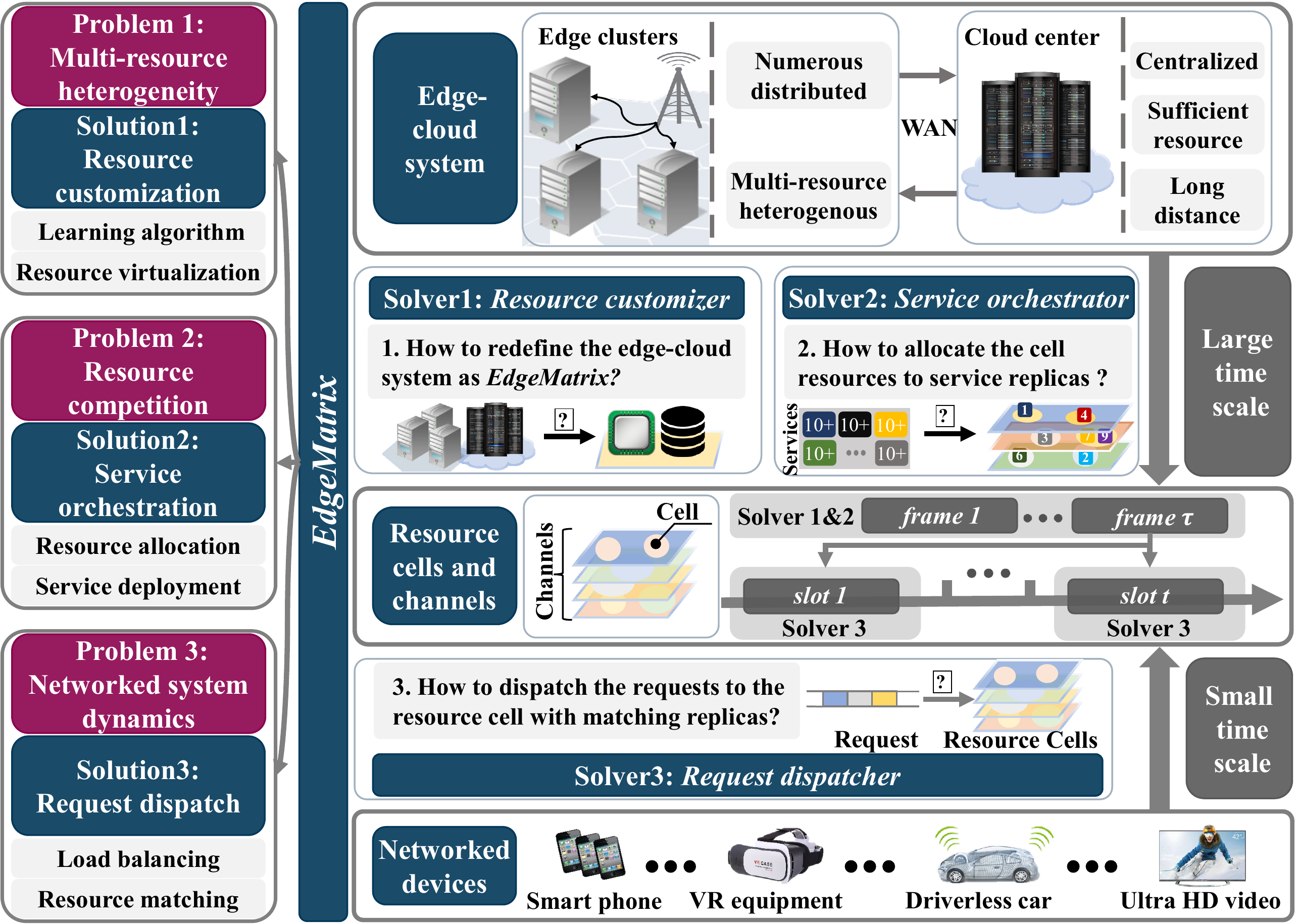}
	\setlength{\abovecaptionskip}{-0.5cm} 
	\caption{Resource customization with joint service orchestration and request dispatch in \textit{EdgeMatrix}.}
	\label{fig:Problem}
	\vspace{-1em}
\end{figure}

\begin{itemize}[leftmargin=*]
\item We design \textit{EdgeMatrix} to redefine heterogeneous physical resources as isolated resources (i.e., resource customization) and solve the problem of joint service orchestration and request dispatch (JSORD).

\item We propose a Networked Multi-agent Actor-Critic (NMAC) algorithm for resource customization with limited neighbor nodes, which provides lightweight models and improves the system's stability through offline centralized training and online distributed execution.


\item We solve the JSORD with multiple types of resources in \textit{EdgeMatrix} based on mixed-integer linear programming (MILP), and significantly reduces the {\color{red}runtime} of the solution by running a multi-task mechanism in parallel.


\item We design a two-time-scale framework for \textit{EdgeMatrix} to coordinate each component, performing resource customization and service orchestration in each frame and request dispatch in each slot as shown in Fig. \ref{fig:Problem}, which outperforms other schemes by real trace-based evaluation.

\end{itemize}


\section{System Model and Problem Statement}
\label{sec:System Model and Problem Statement}

\subsection{Edge-cloud System}
\vspace{-0.3em}



As shown in Fig. \ref{fig:SystemArchitechture}, the edge-cloud system is composed of the network edge and the cloud center. At the edge of the network, there exist massive heterogeneous edge computing nodes, and adjacent nodes in certain regions $\mathcal{D}=\{1,...,D\}$ together form edge clusters to provide closer resources for users. 
The cloud center has sufficient resources, but the centralized deployment approach results in being geographically far from the users. In addition, the edge clusters in each region connect to the cloud center through Wide Area Network (WAN). To be more concise, we (\textit{$\romannumeral1$}) only focus on one region $d \in \mathcal{D}$ at the edge of the network, it also applies to other regions $d' \in \mathcal{D}$, (\textit{$\romannumeral2$}) each edge node can only be connected to a limited number of other nodes, and (\textit{$\romannumeral3$}) assume that each node in the edge cluster has decision-making ability.
\begin{figure}[t]
	\centering
	\includegraphics[width=8.85 cm]{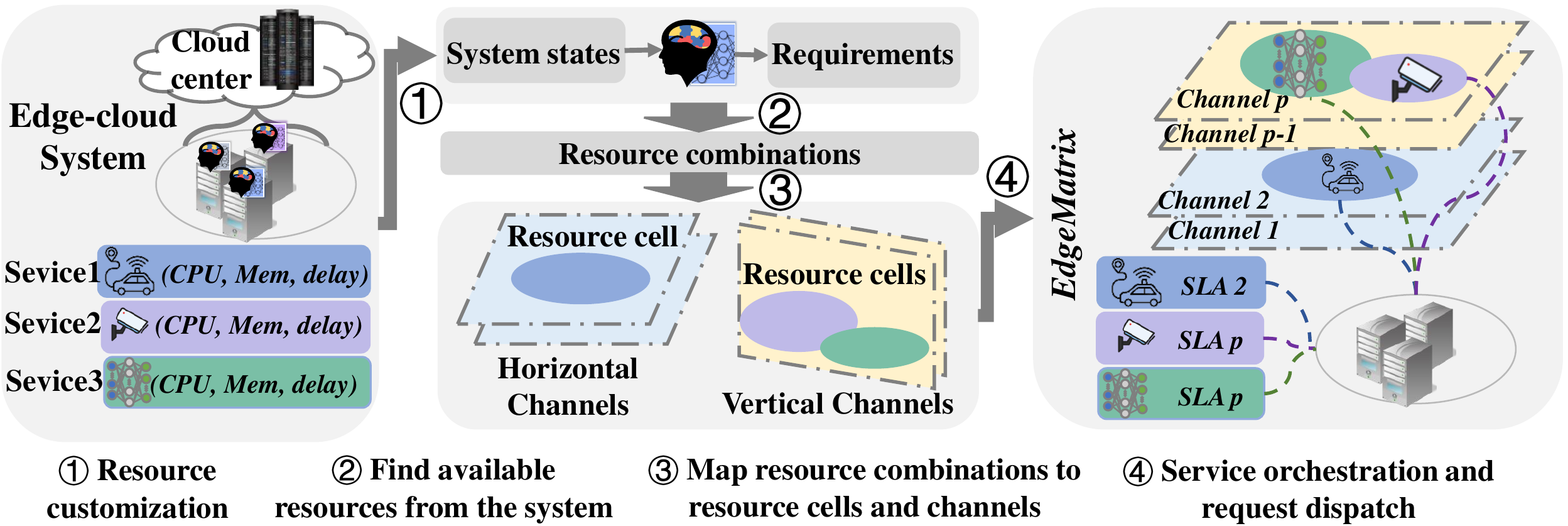}
	\setlength{\abovecaptionskip}{-0.5cm} 
	\caption{Understand \textit{EdgeMatrix} through a case study.}
	\label{fig:CaseStudy}
	\vspace{-1.5em}
\end{figure}

\textbf{\textit{Network edge}}. Geographically dispersed users generate diverse arriving requests over time that have different {\color{red}SLA priorities} $\mathcal{P}=\{1,...,P\}$, and each SLA $p \in \mathcal{P}$ has a service set $\mathcal{L}_{p}=\{1,...,L_{p}\}$. All services with different {\color{red}SLA priority} are denoted by $\mathcal{L}=\mathcal{L}_1\cup \mathcal{L}_2\cup ... \cup \mathcal{L}_P$. For service $l \in \mathcal{L}_p$, we denote the request packet size of each service $l$ as $h_{p,l}$, memory required to load the service $l$ as $r_{p,l}$, the required computing capacity of service $l$ as $w_{p,l}$, the maximum response time of services with $p$ (the lifecycle of service) as $t_{p,l}$, and the execution time of services with $p$ as $o_{p,l}$. 
We denote the network topology of the edge cluster in region $d$ as a graph $G_{d}(\mathcal{V}, \mathcal{E})$, where each $i \in \mathcal{V}$ is the edge node, and $e_{ij} \in \mathcal{E}$ is the link directly connected between node $i$ and node $j$. $\mathcal{N}_{i}=\{j \mid j \in \mathcal{V}, e_{ij} \in \mathcal{E}  \}$ represents the neighborhood where the node $i$ is located, that is, the set of $i$ and its adjacent nodes. The number of edge nodes in cluster $G_{d}$ is denoted as $N$.
Each edge node $i$ has resource capability, we denote the computing capacity as $W_{i}$, the total memory as $R_{i}$, the total bandwidth as $B_{i}$. 


\textbf{\textit{Cloud center}}. The cloud center has sufficient computing and memory resources, but the centralized deployment approach results in a long geographic distance from most users. Therefore, the cloud center is better at handling requests that require a large amount of computing or memory resources but are not latency-sensitive, such as model training. 
We denote the computing capacity owned by the cloud center as $W_{cloud}$, the memory as $R_{cloud}$.



\textbf{\textit{Resource cells and channels}}. \textit{EdgeMatrix} fully collaborates with the advantages of the computing resources between the cloud center and the network edge, customizes a series of resource cells in the edge-cloud system across edge-edge nodes (Horizontal) and edge-cloud nodes (Vertical). Resource cells have a mapping relationship with physical resources in the edge-cloud system, and resource cells are logically isolated with no mutual interference. In addition, we assign corresponding services to the resource cell according to its characteristics (e.g., location, resources). Furthermore, we group resource cells with similar characteristics into one resource channel. The customized services for users with different {\color{red}SLA priorities} are exhibited in the resource channels, which means a channel provides users with a corresponding SLA priority, as shown in Fig. \ref{fig:CaseStudy}. Therefore, we can treat the channels and SLA of services as equivalent to $\mathcal{P}=\{1,...,P\}$. On each channel $p \in \mathcal{P}$, we deploy customized resource cells $\mathcal{M}_{p}=\{1,...,m_{p}\}$ for users according to the SLA of services arriving in the edge cluster. For cell $m \in \mathcal{M}_{p}$, we denote its computing capacity as $W_{p,m}$, and memory size as $R_{p,m}$.


\vspace{-0.7em}
\subsection{Problem Statement}
\label{subsec:problem statement}
\vspace{-0.2em}


The {\color{red}objective} of \textit{EdgeMatrix} is to maximize the overall throughput while reducing SLA violations for various services. To ensure the robustness of \textit{EdgeMatrix}, we adopt a two-time-scale framework in the edge-cloud system to realize resource customization, service orchestration and request dispatch. 

At the large time scale, frame $\tau$, \textit{EdgeMatrix} performs two steps to {\color{red}guarantee the SLA priorities} of different user services: (\textit{$\romannumeral1$}) resource customization, which customizes the resources in the edge-cloud system into resource cells according to the states of the system based on MADRL algorithm, and groups cells with similar characteristics into one resource channel using a clustering algorithm; (\textit{$\romannumeral2$}) service orchestration, which allocates the cell resources to service replicas and then binds the service replicas with allocated physical resources.

\begin{figure}[t]
	\centering
	\includegraphics[width=8.85 cm]{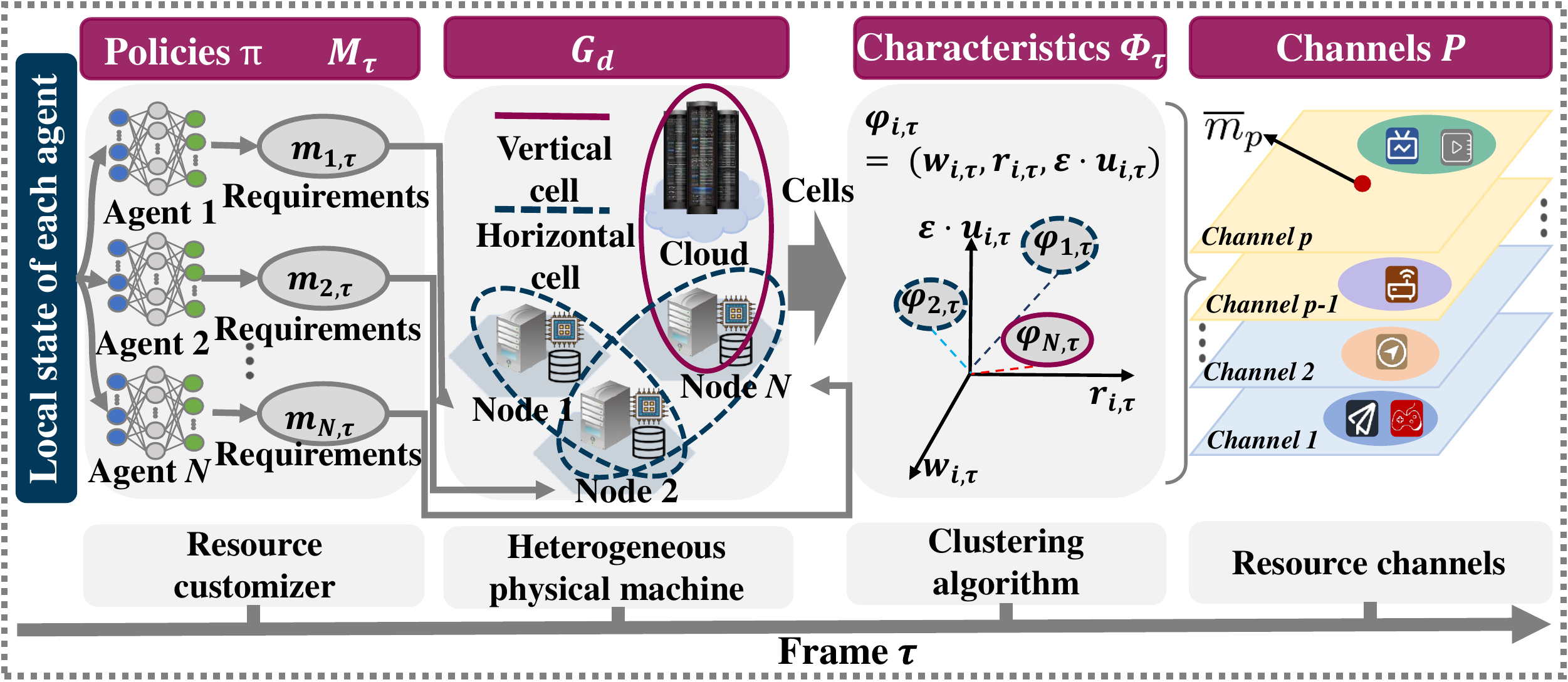}
	\setlength{\abovecaptionskip}{-0.5cm} 
	\caption{The workflow of \textit{resource customizer} at the large time scale.}
	\label{fig:The Workflow of resource customizer}
	\vspace{-1.4em}
\end{figure}

At the small time scale, slot $t$, \textit{EdgeMatrix} performs request dispatch to adapt the networked system dynamics. In \textit{EdgeMatrix}, we implement these three steps by \textit{resource customizer}, \textit{service orchestrator} and \textit{request dispatcher}.

\textbf{\textit{Resource customizer}}. The workflow of \textit{resource customizer} is as shown in Fig. \ref{fig:The Workflow of resource customizer}. We deploy a \textit{resource customizer} agent for every node $i$ in the edge cluster. At frame ${\tau}$, all of the agents need to calculate new resource cell's resource requirements through the observed local state and learned resource customized policy $\pi_{i,\tau}$, denote by $M_{\tau}=\{m_{1,\tau},...,m_{i,\tau},...,m_{N,\tau}\}$. For $m_{i,\tau} \in M_{\tau}$, we denote the memory requirement as $R_{m_{i,\tau}}$ and the computing requirement as $W_{m_{i,\tau}}$. Note that, $M$ is different from $\mathcal{M}$ temporarily, each $m_{i,\tau} \in M_{\tau}$ only represents predicted resource requirements at this moment. The \textit{resource customizer} first obtains the available resources from the neighborhood $\mathcal{N}_i$ of the edge node $i$ where the agent is located according to the requirements of the resources required by $m_{i,\tau}$. When the available resources in the neighborhood can meet the requirements of $m_{i,\tau}$, we call it a horizontal resource cell. \textit{Resource Customizer} will obtain the rest resources if needed from the cloud center, namely the vertical resource cell.

After \textit{resource customizer} finishes customizing the physical resources in the edge-cloud system into resource cells based on the MADRL algorithm, it has to cluster the resource cells to corresponding resource channels. First, the \textit{resource customizer} abstracts the characteristics from each cell, and then it uses a clustering algorithm to group the resource cells with similar characteristics to one resource channel. Finally, SLA priorities are defined for each channel to provide services to users with a corresponding {\color{red}service level}.
Horizontal resource cells have lower transmission latency but limited resources, while vertical resource cells have sufficient resources but higher transmission latency. We then group the cells with similar characteristics into the same channel, and the characteristics of resource cells in \textit{EdgeMatrix} is denoted by $\Phi_{\tau}=\{\varphi_{1,\tau}, ..., \varphi_{i,\tau},...\}$. Specifically, the characteristic of each resource cell $m_{i,\tau}$ is defined by $ \varphi_{i,\tau} = (w_{i,\tau}, r_{i,\tau}, \varepsilon\cdot u_{i,\tau}) $, where $w_{i,\tau}$ and $r_{i,\tau}$ are normalized CPU and memory resources, and $u_{i,\tau}$ is the edge resources proportion. {\color{red}This article mainly focuses on the two resources of CPU and memory because the datasets\cite{Alibabacluster} we are using are computationally intensive services. Note that the larger value of $u_{i,\tau}$ means the lower latency of $m_{i,\tau}$, and the latency is one of the essential factors affecting SLA priority, so we add the consideration of the weighting factor $\varepsilon$ to $u_{i,\tau}$. The larger the value of each item in $\varphi_{i,\tau}$, the better the performance of corresponding item will be in the resource cell $m_{i,\tau}$.} We group the cells $m_{i,\tau} \in M_{\tau}$ based on $\Phi_{\tau}$ into resource channels  $p \in \mathcal{P}$ with corresponding SLA priority using clustering algorithm, i.e., $M_{\tau} \Rightarrow \mathcal{M}_{\mathcal{P}}$. The {\color{red}SLA priority} of each channel is denoted by $ \delta_{p} = \sqrt{w_{\overline{m}_{p}}^{2}+r_{\overline{m}_{p}}^{2}+(\varepsilon\cdot u_{\overline{m}_{p}})^{2}} $, where $\overline{m}_{p}$ is the central point of channel $p$, {\color{red}and SLA priority is proportional to the performance of the resource cell.}

\textbf{\textit{Service orchestrator}}. To make full use of the resources customized by \textit{resource customizer}, the \textit{service orchestrator} needs to appropriately orchestrate service replicas at resource cells on each resource channel. On resource channel $p \in \mathcal{P}$, orchestrating a service replica $l_{p}$ to the cell $m_{p}$, can be denoted as $(l,m)$. We can define all orchestration sets as $\mathcal{S} \subseteq \mathcal{L}_{p} \times \mathcal{M}_{p}$, and describe each single service orchestration as selecting an element from the set. So we can transform the service orchestration problem into set optimization. Note that service orchestration significantly influences request dispatch, and we will explain the relationship between them in the following.

\begin{figure}[t]
	\centering
	\includegraphics[width=8.85 cm]{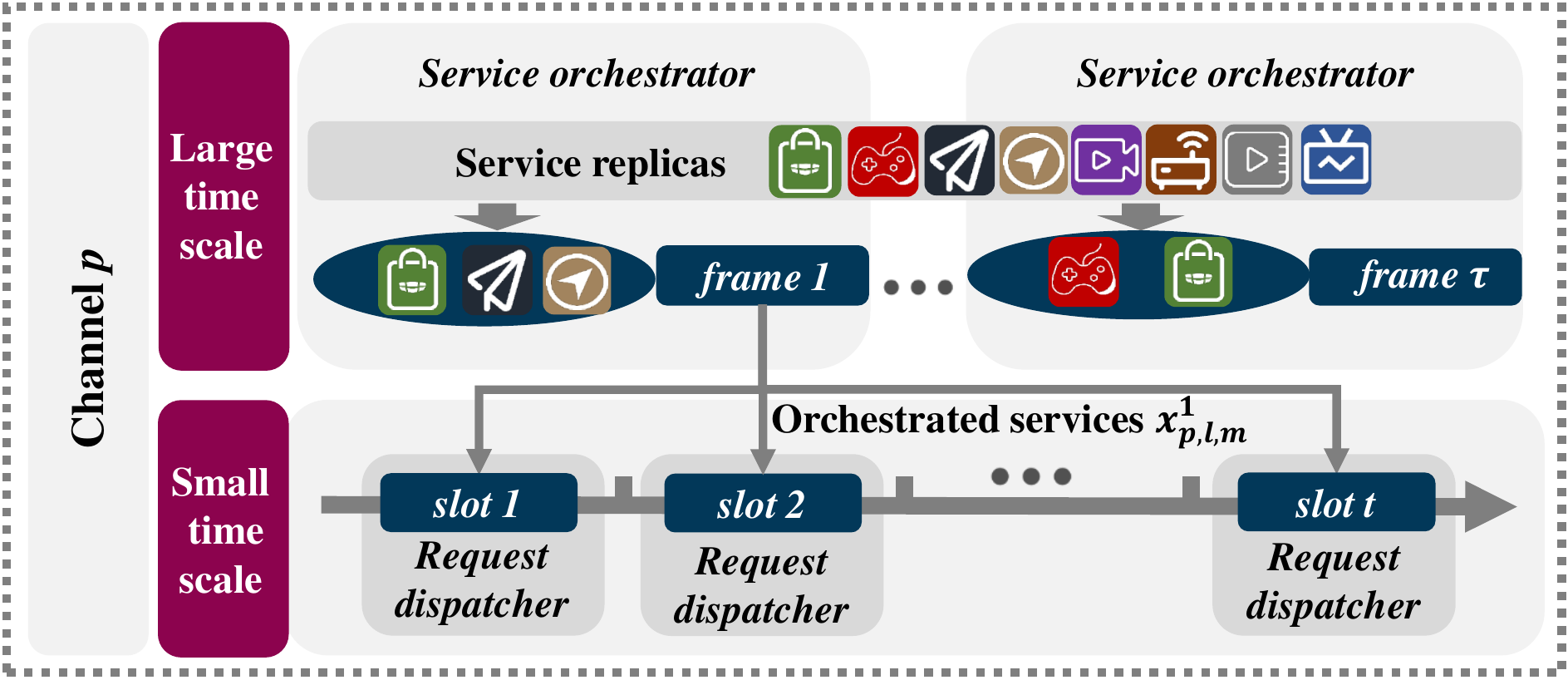}
	\setlength{\abovecaptionskip}{-0.5cm} 
	\caption{The workflow of \textit{service orchestrator} and \textit{request dispatcher} at the two-time-scale framework.}
	\label{fig:Section2_2_JSORD_Problem}
	\vspace{-1.8em}
\end{figure}

\textit{\textbf{Request dispatcher}}. After the resource customization and service orchestration is completed, \textit{request dispatcher} will dispatch the requests that reach the nodes to the resource cells with matching service replicas at the small time scale, slot $t$, as shown in Fig. \ref{fig:Section2_2_JSORD_Problem}. The number of requests from users for service $l \in \mathcal{L}_{p}$ arrived at node $i$ at slot $t$ is $\lambda^{t}_{p,l,i}$, and the average number of requests for a frame $\tau$ is denoted by $\lambda^{\tau}_{p,l,i}$.


To achieve the {\color{red}objective} of \textit{EdgeMatrix}, we have made the following efforts:  (\textit{$\romannumeral1$}) the policy learned by the \textit{resource customizer} reduces SLA violations for various user services (detailed in Sec. \ref{subsec:MADRL for Resource Customization}); (\textit{$\romannumeral2$}) moreover, we jointly consider service orchestration and request dispatch to maximize the overall throughput of the system (i.e., JSORD); and model them together as a mathematical problem because of their strong correlation (a brief introduction in the following, and more details in the Sec. \ref{subsec:Joint Service Orchestration and Request Dispacth}).

Since each channel in $\mathcal{P}$ is similar in terms of handling JSORD, for clarity, one channel $p \in \mathcal{P}$ is used to introduce \textit{EdgeMatrix} in the following discussion. We first set up two decision variables $x$ and $y$, where $x$ is the service orchestration variable and $y$ is the request dispatch variable. More specifically, $x^{\tau}_{p,l,m} \in \left\{0,1\right\}$ is 1 if service $l$ is orchestrated on cell $m$ in frame $\tau$ and 0 otherwise, $y_{p,l,i,m}^{t} \in [0,1]$ represents the probability that a request of service $l$ arrived at edge node $i$ is dispatched to cell $m$ at slot t. For frame $\tau$, we define $y$ as $y_{p,l,i,m}^{\tau} \in [0,1]$. 

We formulate the JSROD as Eq. (1): The object of (1a) is to maximize the number of each channel served requests, $\Psi_{p}=\sum_{l\in\mathcal{L}_{p}}\sum_{i \in \mathcal{V}}\lambda_{p,l,i}\sum_{m\in \mathcal{M}_p}y_{p,l,i,m}$, which equivalent to the system overall throughput because the joint optimization among channels is mutually independent. Constraint (1b) guarantees the request dispatch variable is available. Constraints (1c) and (1d) ensure that each cell's memory and computing capacity can offer the resources required by orchestrated service replicas. {\color{red}Constraint (1e) ensures that $y$ is valid if and only when the service $l$ is orchestrated and won't trigger the SLA}, where $\mathbb{I}_{t_{c,l}-o_{c,l}-t_{i,m} \textgreater 0}$ is the indicator function, indicates the {\color{red}SLA priority} for the requests, and $t_{i,m}$ is the transmission latency between node $i$ and cell $m$. Constraint (1f) is the available range of values.

\vspace{-0.5em}
\begin{gather}
	{\rm max} \ \Psi_{p}, \qquad   \tag{1a} \\
	\textbf{{\rm s.t.}} \sum_{m\in \mathcal{M}_p} y_{p,l,i,m} \leq 1, \quad\qquad\qquad\qquad\qquad\qquad\quad\quad  \tag{1b}\\
	\ \sum_{l\in \mathcal{L}_p} x_{p,l,m} r_{p,l} \leq R_{p,m}, \quad\qquad\qquad\qquad\qquad\   \tag{1c}\\
	\ \sum_{l\in \mathcal{L}_p} \omega_{p,l} \sum_{i \in \mathcal{V}} \lambda_{p,l,i}y_{p,l,i,m} \leq W_{p,m}, \qquad\qquad\quad  \tag{1d}\\
	y_{p,l,i,m} \leq {\rm min}\{x_{p,l,m}, \mathbb{I}_{t_{p,l}-o_{p,l}-t_{i,m} \textgreater 0}\},  \quad\quad   \tag{1e}\\
	\ x\in \{0, 1\}, y \geq 0, \forall p \in \mathcal{P},\ l\in \mathcal{L}_p,\ i \in \mathcal{V},\ m\in \mathcal{M}_p. \tag{1f}
\end{gather}


\vspace{0.5em}
\section{Algorithm Design}
\label{sec:Algorithm and System Design}


\subsection{MADRL for Resource Customization}
\vspace{-0.3em}
\label{subsec:MADRL for Resource Customization}
The \textit{resource customizer} agent deployed on each node can calculate new resource cells' resource requirements for \textit{EdgeMatrix} based on local state and learned policy. Its {\color{red}objective} is to provide customized resource cells for user services with different {\color{red}SLA priority} under (\textit{$\romannumeral1$}) multi-resource heterogeneous edge nodes and (\textit{$\romannumeral2$}) dynamically changing service requests, which reduces SLA violations for various user services.

With the development of artificial intelligence, especially reinforcement learning (RL)\cite{sutton2018} represented by DQN\cite{mnih2015}, DDPG\cite{timothy2016ddpg} and A3C\cite{mnih2016}, game control and robot control are well performed by RL. Due to the large number of computational nodes distributed in the edge-cloud system, the direct implementation of these algorithms will cause a high-dimensional action space or non-stationary environment\cite{lowe2017multi,chu2019}. Therefore, we introduce the MADRL algorithm to enable each decision-capable edge node in the system to customize the resources in its network neighborhood into resource cells based on the changing system state. To learn practical resource customization policies in complex networked environments, we must consider (\textit{$\romannumeral1$}) the impact of algorithm training on the robustness of the networked system, (\textit{$\romannumeral2$}) the unsuitability for edge nodes with limited computational power to deploy large models, and (\textit{$\romannumeral3$}) the high-dimensional action space in the decision making of the networked system. Therefore, we adopt an algorithmic framework of offline centralized training and online distributed execution with a continuous action space.

\subsubsection{Markov Game Formulation}

Since the edge cluster of each region $d \in \mathcal{D}$ in \textit{EdgeMatrix} is a graph $G_{d}$, a multi-agent Markov Decision Process (MDP) can be formed as $\rho=(G_{d},\{\hat{\mathcal{S}_{i}},\hat{\mathcal{A}_{i}}\}_{i \in \mathcal{V}},\hat{\mathcal{P}}, \{\hat{\mathcal{R}_{i}}\}_{i \in \mathcal{V}})$. We denote the resource customizer agent on each edge node as $i \in \mathcal{V}$. More details are introduced according to $\rho$ in the following. 
%
\vspace{-0.3em}

\textbf{\textit{State space $\hat{\mathcal{S}}$}}. At frame $\tau$, the local state space observed by the agent on edge node $i$ is $\hat{s}_{i,\tau}$, which contains: (\textit{$\romannumeral1$}) the number and kinds of requests $(\lambda_{1,1,i}^{\tau}, ..., \lambda_{p,l,i}^{\tau})_{p\in\mathcal{P},l\in\mathcal{L}_{p}}$; (\textit{$\romannumeral2$}) the resource requirements and delay demand of requests arrived at node $i$; (\textit{$\romannumeral3$}) the CPU, memory and edge resources proportion of the existing cells created by agent $i$ in the system; (\textit{$\romannumeral4$}) the available resources of edge nodes where agent $i$ is located, $\mathcal{N}_{i}$. We simply consider the global observation for the training critic as the ensembles of all agents' state, $\bm{\hat{s}}_{\tau}$.

\textbf{\textit{Action space $\hat{\mathcal{A}}$}}. We define the action space of all agents in the edge cluster as a joint action space $\hat{\mathcal{A}}=\{\hat{\mathcal{A}}_{1},...,\hat{\mathcal{A}}_{i},...,\hat{\mathcal{A}}_{N}\}_{i \in \mathcal{V}}$ , where $\hat{\mathcal{A}}_{i} \in \hat{\mathcal{A}}$ represents the action space of agent $i$. At frame $\tau$, agent $i$ predict the action $\hat{a}_{i,\tau}$ according to the observed local state space $\hat{s}_{i,\tau}$ and policy $\pi_{i,\tau}$. Specifically, $\hat{a}_{i,\tau}$ indicates the size of the resource that the agent $i$ predicts allocate to cell $m_{i,\tau}$, i.e., $(W_{m_{i,\tau}}, R_{m_{i,\tau}})$, where both of them are continuous variables with a value range of $[0, 1]$. The actual resource size is $(\alpha \cdot W_{m_{i,\tau}}, \beta \cdot R_{m_{i,\tau}})$, where $\alpha$ and $\beta$ are the upper limits of the cell's resources.

\textbf{\textit{Reward function $\hat{\mathcal{R}}$}}. Agent $i$ inputs the observed local state space $\hat{s}_{i,\tau}$ and selected action $\hat{a}_{i,\tau}$ at frame $\tau$ into the reward function $\hat{\mathcal{R}}$ to get an immediate reward $\hat{r}_{i,\tau}$. To learn how to improve the overall throughput of the system while reducing SLA violations for various user services, we comprehensively consider service throughput and SLA priority to help the agent learn this ability in an environment where multi-agents coordinate with each other. The reward function can be formulated as $\hat{r}_{i,\tau}=\sum_{p\in\mathcal{P}}\delta_{p}\sum_{l\in\mathcal{L}_{p}}\Psi'_{l,i,\tau}$, where $\Psi'_{l,i,\tau}=\Psi_{l,i,\tau}/\lambda_{p,l,i}^{\tau}$ is the throughput rate of service $l$ arrived at node $i$ at $\tau$. {\color{red}$\delta_{p}$ indicates the weight of the services with SLA priority $p$, and we verified the necessity of this setting by adjusting $\mathcal{\varepsilon}$ in Sec. \ref{subsec:Setting of Key Parameters}}.

\textbf{\textit{State transition function $\hat{\mathcal{P}}$}}. Note that we use a deterministic policy, and the state transition function is denoted as $\hat{\mathcal{P}}(\hat{\mathcal{S}}' \mid \hat{\mathcal{S}},\hat{\mathcal{A}}_{1},...\hat{\mathcal{A}}_{N}): \hat{\mathcal{S}} \times \hat{\mathcal{A}}_{1} \times \ldots \times \hat{\mathcal{A}}_{N} \mapsto \hat{\mathcal{S}}'$.
\begin{figure}[t]
	\centering
	\includegraphics[width=8.85 cm]{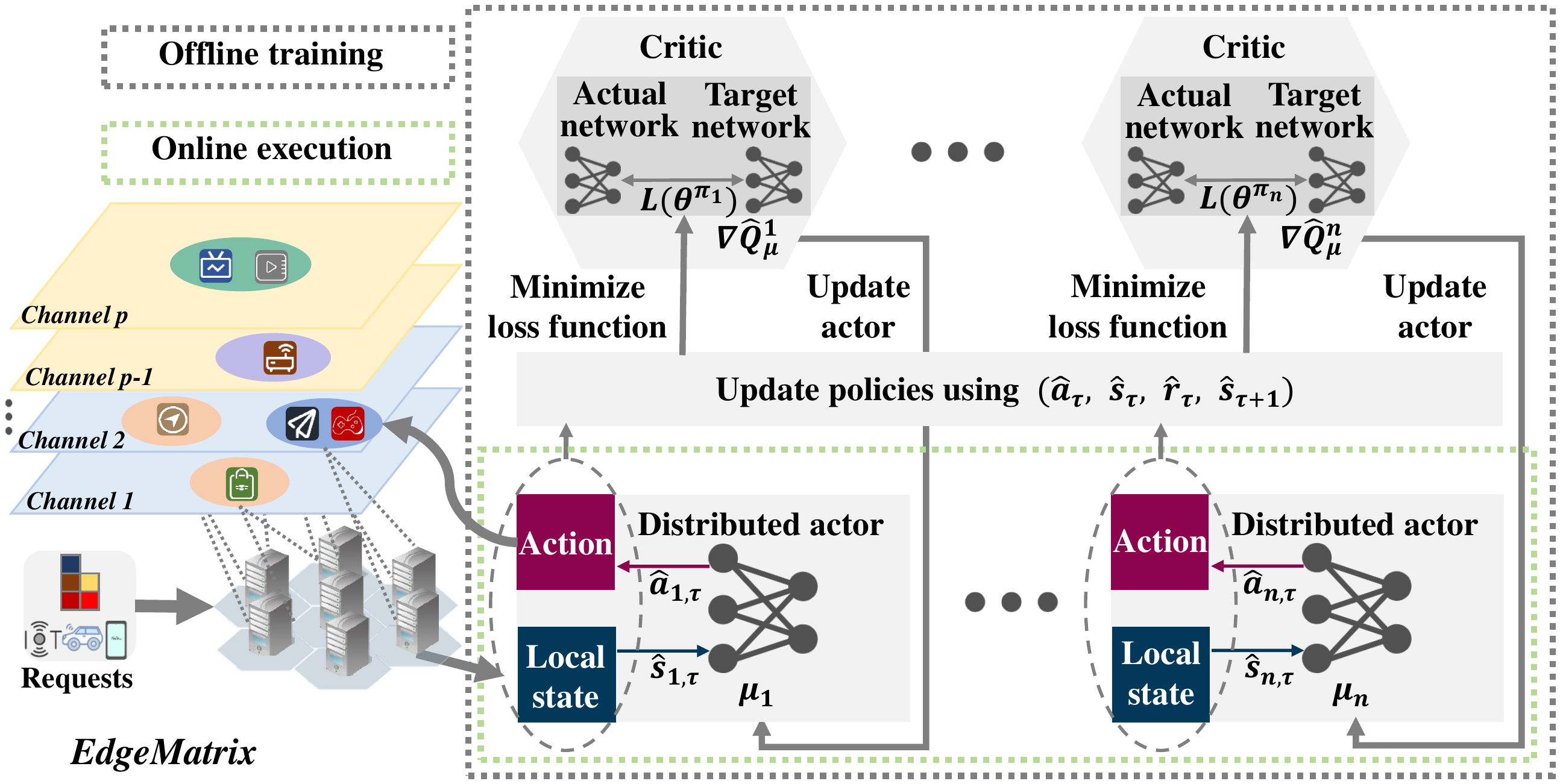}
	\setlength{\abovecaptionskip}{-0.5cm} 
	\caption{Networked multi-agent actor-critic.}
	\label{fig:NMAC}
	\vspace{-1.9em}
\end{figure}

\vspace{-0.1em}
\subsubsection{Networked Multi-agent Actor-Critic}
\label{subsubsec:Coordinated Multi-agent Actor-Critic}
In the networked environment of the edge-cloud system, the main problems in designing the MADRL algorithm are: (\textit{$\romannumeral1$}) we need to minimize the requirements of resources for agents' decision-making because the computing nodes at the edge of the network only have limited resources; (\textit{$\romannumeral2$}) reducing SLA violations is the aim of our work, so the training or execution process should not affect the stability and security of the networked system. 


To better deal with the above problems, we proposed the \textit{Networked Multi-agent Actor-Critic (NMAC)} algorithm in the multi-agent coordinate environment of edge-cloud system, as shown in Fig. \ref{fig:NMAC}, (\textit{$\romannumeral1$}) centralized critic, which can guide each actor to learn an effective policy according to global observation with extra information during training; (\textit{$\romannumeral2$}) distributed actor, each actor's input during training and execution is local state, so the actor can seamlessly switch between the two phases.
\begin{itemize}[leftmargin=*]

\vspace{-0.3em}
\item \textbf{\textit{Centralized critic}}. During the training process, we equip each agent with a critic to train the actor. For agent $i$, the critic is implemented based on the centralized action-value function $\hat{Q}\left(\bm{\hat{s}}_{\tau}, \bm{\hat{a}}_{\tau} \mid \theta^{\pi_{i}}\right)$, which represents the expected discounted cumulated reward of frame $\tau$ starting from state-action pairs $(\bm{\hat{s}}_{\tau}, \bm{\hat{a}}_{\tau})$ according to the policy $\pi_{i}$, $\bm{\hat{a}}_{\tau}=(...,\hat{a}_{i,\tau},...)$. The action-value function can be represent as $\hat{Q}\left(\bm{\hat{s}}_{\tau}, \bm{\hat{a}}_{\tau} \mid \theta^{\pi_{i}}\right)=\mathbb{E}_{\pi_{i}}\left[R_{i,\tau}\right]$, where $R_{i,\tau}=\hat{r}_{i,\tau}+\sum_{\tau'=\tau+1}^{\mathcal{T}}\gamma^{(\tau'-\tau)}\hat{r}_{i,\tau+1}$. Thus, the centralized action-value function can be obtained from the Bellman Equation:
\vspace{-0.5em}
\begin{equation}
	\hat{Q}\left(\hat{\bm{s}}_{\tau}, \hat{\bm{a}}_{\tau} \mid \theta^{\pi_{i}}\right)=\hat{r}_{i,\tau}+\gamma \max \limits_{\hat{\bm{a}}_{\tau+1}} \hat{Q}\left(\hat{\bm{s}}_{\tau+1}, \hat{\bm{a}}_{\tau+1} \mid \theta^{\pi_{i}}\right),\tag{2}
\end{equation}
where $\theta^{\pi_{i}}$ is the parameter of policy $\pi_{i} \in \Pi$, $\Pi=\{\pi_{1},...,\pi_{N}\}$, and the optimization function can be derived as a loss function between actual critic network $\hat{Q}_{i,\tau}$ and target critic network $\hat{G}_{i,\tau}$, 
\begin{equation}
	\label{equation:Critic update function}
	\begin{array}{l}
		L\left(\theta^{\pi_{i}}\right)=\mathbb{E}\left[\left(\hat{Q}\left(\hat{\bm{s}}_{\tau}, \hat{\bm{a}}_{\tau} \mid \theta^{\pi_{i}}\right) - \hat{G}_{i,\tau}\right)^{2}\right],\\
		\text { where } \quad \hat{G}_{i,\tau}=\hat{Q}^{\prime}\left(\hat{\bm{s}}_{\tau}, \hat{\bm{a}}_{\tau} \mid \theta^{\pi'_{i}}\right).\tag{3}
	\end{array}
\end{equation}

\item \textbf{\textit{Distributed actor}}. For each agent, the actor network learns a deterministic policy $\mu_{i}$ to maximize the cumulative reward, i.e., $J=\mathbb{E}_{\mu_{i}}\left[R_{i,\tau}\right]$. We update the parameters $\theta^{\mu_{i}}$ through optimizing policy gradient:
\begin{multline}
	\label{equation:Actor update function}
	\nabla_{\theta^{\mu_{i}}} J\left(\mu_{i}\right)=\\
	\mathbb{E}\left[\nabla_{\theta^{\mu_{i}}} \log \mu_{i}\left(\hat{a}_{i,\tau} \mid \hat{s}_{i,\tau}\right) \nabla_{a_{i}}\hat{Q}\left(\hat{s}_{i,\tau}, \hat{a}_{i,\tau}\mid\theta^{\mu_{i}}\right)\right]. \tag{4}
\end{multline}

\end{itemize}

Especially, \textit{NMAC} implements an offline training and online execution framework: (\textit{$\romannumeral1$}) offline-training, can avoid that the training process may have a negative impact on the networked system stability; (\textit{$\romannumeral2$}) online-execution, only requires the actor-network to predict the action and the learned policy only uses local state, which significantly reduces the resources consumed by the agent compared to the training phase.

\vspace{-0.3em}
\vspace{-0.3em}
\subsection{Joint Service Orchestration and Request Dispacth}
\label{subsec:Joint Service Orchestration and Request Dispacth}
\vspace{-0.4em}

Since service orchestration significantly impacts request dispatch, we together consider them as a joint optimization problem, i.e., JSORD. Specifically, (\textit{$\romannumeral1$}) at the large time scale, JSORD has to orchestrate the appropriate services for each cell in \textit{EdgeMatrix} based on the system state, and (\textit{$\romannumeral2$}) at the small time scale, JSORD has to dispatch the requests arriving at each node of the system to resource cells. Considering the system with multiple types of constraints, (e.g., computation, memory, communication, and latency requirements), we solve JSORD based on MILP. However, the widely distributed edge nodes and the variety of services in the system make the {\color{red}runtime} of the solution unacceptable for users.



{\color{red}In \textit{EdgeMatrix}, each channel serves a specific class of services with the same SLA priority. The resources used during service orchestration and request dispatch are the cells of the channel, so the channels are independent of each other. Therefore, we execute JSORD independently in parallel on each channel, which significantly reduces the {\color{red}runtime}.}



\begin{figure}[t]
	\centering
	\includegraphics[width=8.85 cm]{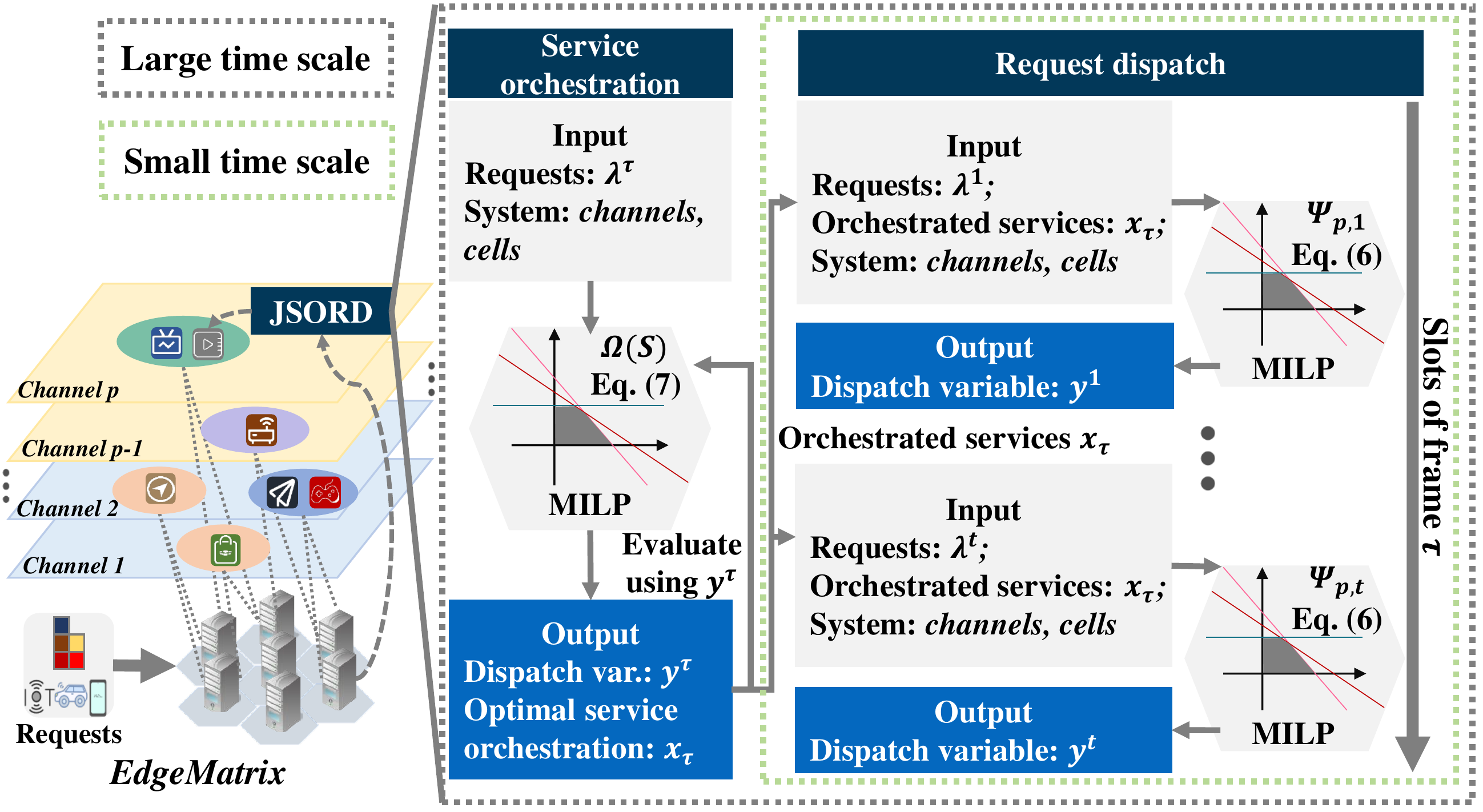}
	\setlength{\abovecaptionskip}{-0.5cm} 
	\caption{Joint service orchestration and request dispatch.}
	\label{fig:JSORD}
	\vspace{-1.4em}
\end{figure}

We model the joint optimization problem of service orchestration and request dispatch as Eq. (1) at the end of section \ref{sec:System Model and Problem Statement}. More specifically, (\textit{$\romannumeral1$}) at the beginning of frame $\tau$, we calculate the optimal service orchestration $x_{p,l}^{\tau}$ based on Eq. (1) by the predicted request dispatching probability $y_{p,l,i,m}^{\tau}$ and the request demand $\lambda_{p,l,i}^{\tau}$; (\textit{$\romannumeral2$}) then at the beginning of slot $t$, we calculate the request dispatch variable $y_{p,l,i,m}^{t}$ with the current request demand $\lambda_{p,l,i}^{t}$ and the orchestrated service $x_{p,l}^{\tau}$ to solve Eq. (1). At the large time scale, note that although the dispatch variable $y_{p,l,i,m}^{\tau}$ is predicted according to the demand $\lambda_{p,l,i}^{\tau}$, $y_{p,l,i,m}^{\tau}$ is used to evaluate Eq. (1a) under the given orchestrated services $x_{p,l}^{\tau}$ rather than request dispatch.

\subsubsection{Solvability Analysis}
We first analyze the solvability of the joint optimization Eq. (1) and then consider the special case of Eq. (1), where the resource cell and the service are homogeneous, with constraints (1d) ignored: 
\begin{gather}
	{\rm max} \ \Psi_{p},  \qquad\qquad\quad \tag{5a}\\
	\textbf{{\rm s.t.}} \ (1b),(1e),(1f), \qquad\qquad\qquad\quad\tag{5b}\\
	\sum_{l\in \mathcal{L}_p} x_{p,l,m} \leq R_{p, m}. \qquad\quad\quad \tag{5c}
\end{gather}
\vspace{-0.6em}

The joint optimization Eq. (1) can be simplified to the 2-Disjointed Set Cover Problem, i.e., Eq. (5), which is proved to be NP-complete\cite{cardei2005}. The special case of the joint optimization, Eq. (1) is NP-hard, which means that the joint optimization problem of service orchestration and request dispatch is also NP-Hard in the general case.

To describe the solution process of Eq. (1) more clearly, we discuss the deformations of Eq. (1) and the solution processes at the two-time-scale framework in the following, i.e., service orchestration and request dispatch, as shown in Fig. \ref{fig:JSORD}.

\subsubsection{Requst Dispatch}
\label{subsubsec:GNN-based System State Encoding}

At slot $t$, the \textit{service orchestrator} has already orchestrated the services on each resource cell, which means we solve the request dispatch problem under the situation that service orchestration variable $x_{p,l}^{\tau}$ is known. Thus, the joint Eq. (1) can be simplified to linear programming, i.e., Eq. (6), which means we can get the probability $y_{p,l,i,m}^{t}$ when dispatching a request of service $l$ arriving at the edge node $i$ to cell $m$. (If the request is successfully dispatched with our constraints (6b)-(6d), we can serve it.)
\begin{gather}
	\label{equation:6}
	{\rm max} \ \Psi_{p},  \qquad\qquad  \tag{6a}\\
	\textbf{{\rm s.t.}} (1b), (1d), (1e),(1f), \qquad\qquad  \tag{6b}\\
	\qquad\  y_{p,l,i,m} \leq \mathbb{I}_{(l,m)\in S}, \qquad\quad\   \tag{6c}\\
	\qquad\  y_{p,l,i,m} \in [0, 1]. \qquad\qquad\quad \tag{6d}
\end{gather}

\subsubsection{Approximation Algorithm for Service Orchestration}

Service orchestration problem can be transformed to a set optimization problem in the Sec. \ref{subsec:problem statement}. On resource channel $p \in \mathcal{P}$, orchestrating a service replica $l_{p}$ to the cell $m_{p}$ can be denoted as $(l,m)$. We can define all orchestration sets as $\mathcal{S} \subseteq \mathcal{L}_{p} \times \mathcal{M}_{p}$, and each single service orchestration can be described as selecting an element from the set. Let $\Omega\left(\mathcal{S}\right)$ denote the optimal {\color{red}objective} value of Eq. (1) under a fixed set $\mathcal{S}$ of orchestrated services and a fixed dispatch variable 
\begin{algorithm}
	\caption{Solve JSORD Based on Submodular Function Maximization}
	\label{alg:JSORD}
	\small \KwIn{Input parameters of Eq. (1)\;} 
	\KwOut{Service orchestration variable $x^{\tau}_{p,l}$ and requests dispatch variable $y_{p,l,i,m}^{\tau}$\;}
	
	Initialize  $frame =\tau,\ channel=p$, $\mathcal{S} = \emptyset $, $T = \{e | e \in (\mathcal{L}_p\times \mathcal{M}_p) \setminus \mathcal{S}$, $\mathcal{S}\cup\{e\}$ satisfies constraints of (\ref{equation:7})$\}$ \;
    \While{$ T \neq \emptyset $}  
    	{  
    		$e^* = $ the element $e$ in $T$ that get the maximum value of \textit{\textbf{$\Omega(\mathcal{S}\cup\{e\}$}}\;
    		$\mathcal{S} = \mathcal{S}\cup \{e^*\}$\;
    		$T = \{e | e \in (\mathcal{L}_p\times \mathcal{M}_p) \setminus \mathcal{S}$, $\mathcal{S}\cup\{e\}$ satisfies constraints of (\ref{equation:7})\}\;
    	}
	Convert $\mathcal{S}$ to its vector representation $x^{\tau}_{p,l}$\;
	Compute $y_{p,l,i,m}^{\tau} = \{..., y_{p,l,i,m}^{t}, ...\}$ using Eq. (\ref{equation:6}) based on orchestrated services $x^{\tau}_{p,l}$\;
	\For{slot $t=0,1,2,...,$}
	{
		Execution request dispatch with dispatching variable $y_{p,l,i,m}^{t}$ at the small time scale.
	}

\end{algorithm}

$x$, $(l,m) \in \mathcal{S}$ if and only if $x_{l,m}=1$. This can be calculated by solving the request dispatch problem (see Eq. (6)), and then we can rewrite the problem as:

\vspace{-0.8em}
\begin{gather}
	\label{equation:7}
	{\rm max} \ \Omega(\mathcal{S}), \quad\qquad\  \tag{7a} \\
	\textbf{{\rm s.t.}} \sum_{l:(l, m)\in \mathcal{S}}r_{p,l} \leq R_{p,m},    \qquad \tag{7b}\\
	\qquad \mathcal{S}\subseteq \mathcal{L}_p\times \mathcal{M}_p. \qquad\ \  \tag{7c}
\end{gather}
In summary, the overall training and scheduling process of \textit{EdgeMatrix} is given in Algorithm \ref{alg:OverallAlgorithm}.


\section{Implementation}
\label{sec:System Design and Implementation}

\subsection{Edge-cloud System Setup}
\label{subsec:End-cloud-edge Computing System Setup}
\vspace{-0.3em}


\textbf{\textit{Edge cluster and cloud center}}. In the simulation experiment, we set up $10$ edge nodes and $6$ SLA Allocation/Retention Priority (ARP) services\cite{3GPPwhitepaper} in the edge cluster of a region from the perspective of the service providers. Each property of the edge node $i \in \mathcal{V}$ is set to $W_{i}=\left[2,4\right]$ vCPUs, $R_{i}=\left[100, 200\right]$ GB and $B_{i}=\{125,12.5\}$ Mbps. We assume that the computing capacity $W_{cloud}$ and memory capacity $R_{cloud}$ of cloud center are always sufficient, and the connection between the cloud center and the edge cluster is reliable, so the transmission delay from the edge node to the cloud center is set to a constant $L_{edge}^{cloud}=10 ms$.


\textbf{\textit{Resource cells and channels}}. By default, at each frame $\tau$, the number of resource cells that each agent can maintain is $\sum_{p\in \mathcal{P}}m_{p,i} \geq 1$. The requirements of each resource cell are predicted by the continuous action $\hat{a}_{i,\tau}=\{W_{m_{i,\tau}}, R_{m_{i,\tau}}\}$. These two numbers are floats between $[0,1]$, and their true resource requirements are multiplied by the scalar in the system, i.e., $(2\ vCPUs, 500\ MB)$. These resource cells are classified to the corresponding resource channels according to the cell characteristics $\Phi$. There are $[2,4]$ kinds of service on each channel, and the number of resource channels is $6$.

\begin{algorithm}
	\caption{The Overall Algorithm of \textit{EdgeMatrix}}
	\label{alg:OverallAlgorithm}
	\small Initialize the system environment and training parameters\;
	Get the system observation $\hat{s}_{0}$\;
	\For{frame $\tau=0,1,2,...,$}
	{
		Get the actions of each agent $\hat{a}_{\tau}=\left(\hat{a}_{1,\tau},...,\hat{a}_{i,\tau},...\right)$\;
		Resource Customizer execution actions\;
		\For{channel $p=0,1,2,...,$}
		{
			Solve \textbf{JSORD} based on Algorithm\ref{alg:JSORD} in parallel on each channel $p\in\mathcal{P}$ \;
		}
		Get reward $\hat{r}_{\tau}$ and next observation $\hat{s}_{\tau+1}$\;
		Store [$\hat{s}_{\tau},\hat{a}_{\tau},\hat{r}_{\tau},\hat{s}_{\tau+1}$] for updating neural network\;
		\If{frame \% update rate $==0$}
		{
			Update the parameters of actor ($\theta^{\mu}$) and critic($\theta^{\pi}$) using Eq. (\ref{equation:Critic update function}) and Eq. (\ref{equation:Actor update function})\;
			Save models.
		}
	}
\end{algorithm}

\textbf{\textit{Service and request}}. Our system's data value range and request distribution are based on the \textbf{Alibaba Cluster Trace}\cite{Alibabacluster} to ensure \textit{EdgeMatrix} has effective performance in the real environment. However, since the dataset does not reflect the delay characteristics, we refer to ETSI's white paper\cite{3GPPwhitepaper} to determine the delay data range of requests.

\vspace{-0.5em}
\subsection{Training Settings}
\label{subsec:Training Settings}
\vspace{-0.3em}

We implemented Algorithm 2 based on python 3.6 with the following detailed setup. Each NMAC agent consists of a critic network and an actor policy network, training the network with the fixed learning rate of $\eta=0.01$ and the reward discount factor of $\gamma=0.95$. The critic network is a three-layer fully-connected neural network (FCNN) with 64 neurons per layer, in which the activation function in the first two hidden layers is relu, and there is no activation function in the output layer. The actor policy network is also a three-layer FCNN with 64 neurons per layer, in which the activation function is relu in the first two layers, and the output layer is sigmoid to ensure that the output is in the valid range. The linear program solver in JSORD uses linprog function of the SciPy library.


\section{Performance Evaluation}
\label{sec:Performance Evaluation}



Under the {\color{red}objective} of maximizing the overall throughput while reducing SLA violations for various services, we have verified the effectiveness of \textit{EdgeMatrix} on the three challenges of multi-resource heterogeneity, resource competition, and networked system dynamics. More specifically, we consider three metrics: (\textit{$\romannumeral1$}) the reward used to evaluate the comprehensive performance of \textit{EdgeMatrix} when training the algorithm; (\textit{$\romannumeral2$}) to avoid system throughput $\Psi \rightarrow \infty$, we use the overall throughput rate to verify the system performance; (\textit{$\romannumeral3$}) percentage of requests served by each channel to total requests in \textit{EdgeMatrix} used to reflect \textit{EdgeMatrix}'s {\color{red}assurance of SLA priority}. In our experiments, the baselines include (\textit{$\romannumeral1$}) MADDPG, a general MADRL algorithm; (\textit{$\romannumeral2$}) MA2C, a state-of-art MADRL algorithm applied in the networked system.

\subsection{Setting of Key Parameters}
\label{subsec:Setting of Key Parameters}

\begin{figure}[t]
	\centering
	\includegraphics[width=8.85 cm]{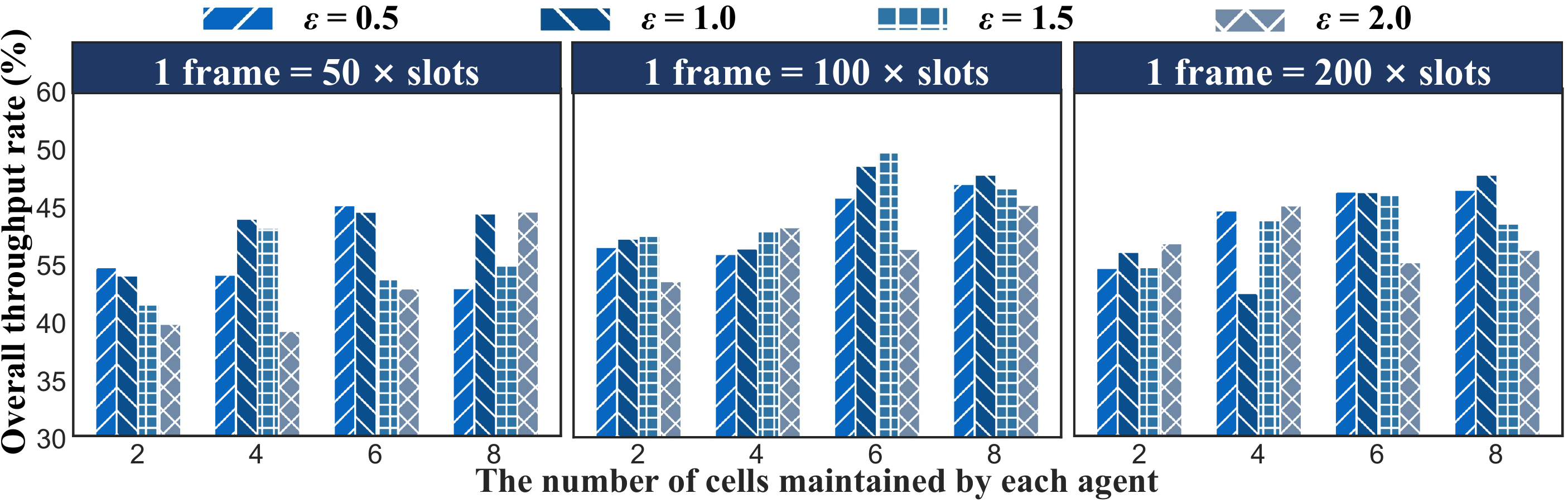}
	\vspace{-2em}
	
	\caption{\textit{EdgeMatrix}'s training performance under various parameters.}
	\label{fig:ParametersIdentify}
	\vspace{-1.8em}
	\hspace{-0.4em}
\end{figure}

We gradually determine several essential parameters during training, as shown in Fig. \ref{fig:ParametersIdentify}. The frequency of service orchestration has a significant impact on the training performance of \textit{EdgeMatrix}, i.e., frequent or occasional service orchestration cannot get better performance. (\textit{$\romannumeral1$}) For \textit{frequent orchestration}, the cell must reload replicas of the service at each service orchestration, making some requests time out during the waiting process; (\textit{$\romannumeral2$}) for \textit{occasional orchestration}, the requests of the networked system are dynamically changing, and the services orchestrated in the system need to be appropriately adjusted. Therefore, we set 100 slots at each frame. The system performance will perform better as the number of cells maintained by each node increases. However, the node maintains cells requires additional resource costs, so we set the number of cells maintained by each node to 6. As one of the essential characters of SLA priority, the edge resources proportion determines the delay guarantee of cells. The large weight of the edge resources proportion causes the impact of core network resources negligible, so we set it as $\mathcal{\varepsilon}=1.5$.
\subsection{Learning Ability of \textit{EdgeMatrix}}


\begin{figure}[t]
	\centering
	\includegraphics[width=8.85 cm]{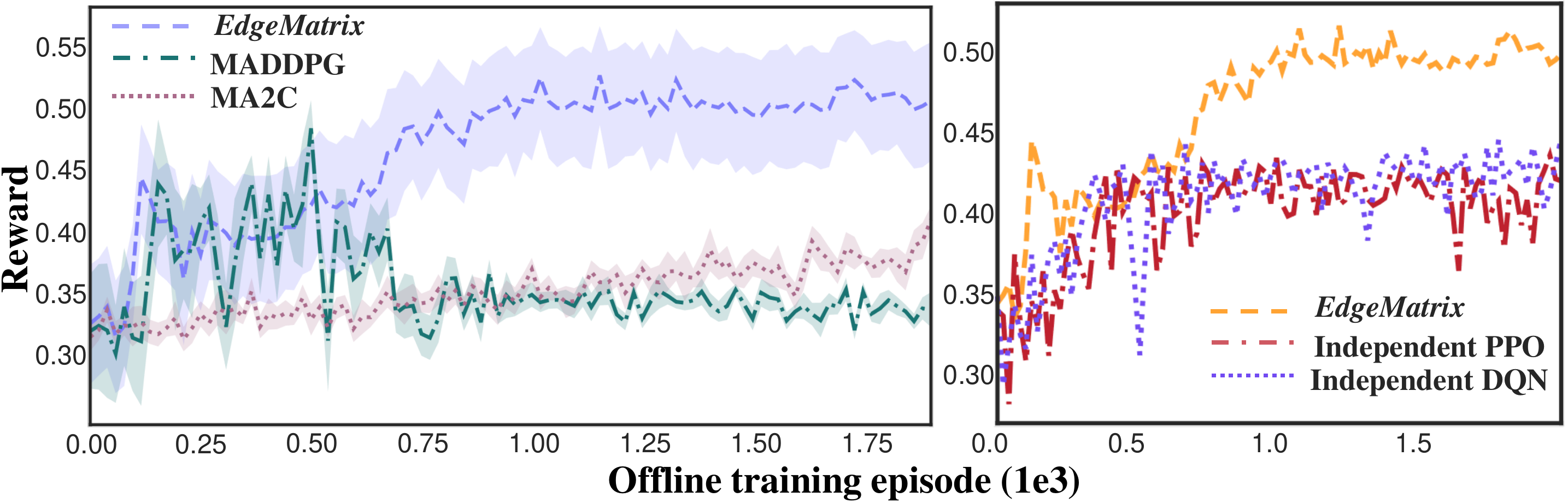}
	\vspace{-2em}
	\caption{Learning ability of \textit{EdgeMatrix} compared with (a) the two baselines, MADDPG and MA2C and (b) independent PPO and independent DQN.}
	\label{fig:LearningAbility}
	\vspace{-1.6em}
\end{figure}

To verify the learning performance of \textit{EdgeMatrix}, we first compare the training performance of the other two baselines, proving that \textit{EdgeMatrix} has the feasibility of convergence and has effective learning capabilities. {\color{red}Then we compare the performance of the algorithm \textit{EdgeMatrix}, independent DQN\cite{mnih2015}, and independent PPO\cite{schulman2017}} to demonstrate the effectiveness of \textit{EdgeMatrix} among 10 edge nodes.

As shown in Fig. \ref{fig:LearningAbility}(a), the three algorithms at the first 100 training episodes are in the exploration stage with random policy, i.e., they do not learn policies and only collect training data, and their rewards are at the same level. The rewards of \textit{EdgeMatrix} and MADDPG begin to rise sharply after the 100th episode and maintain a flattened period. It is because the algorithm has learned some effects from the data collected in the first 100 episodes. However, the learning performance is flattening along with the state of the dynamic networked system. After the 1000th episode, both \textit{EdgeMatrix} and MADDPG tend to converge. In particular, the performance of \textit{EdgeMatrix} is improved by about 60\%, but MADDPG captures almost no knowledge among the data. Moreover, there is no significant impact on the performance of MA2C, which also implies that the algorithm gains little benefit from training based on sample data. Fig. \ref{fig:LearningAbility}(b) demonstrates that a simple implementation of independent agent in the multi-agent environment is not {\color{red}excellent} due to the non-stationary problem.

\subsection{Practicability of \textit{EdgeMatrix}}

\begin{figure}[t]
	\centering
	\includegraphics[width=8.85 cm]{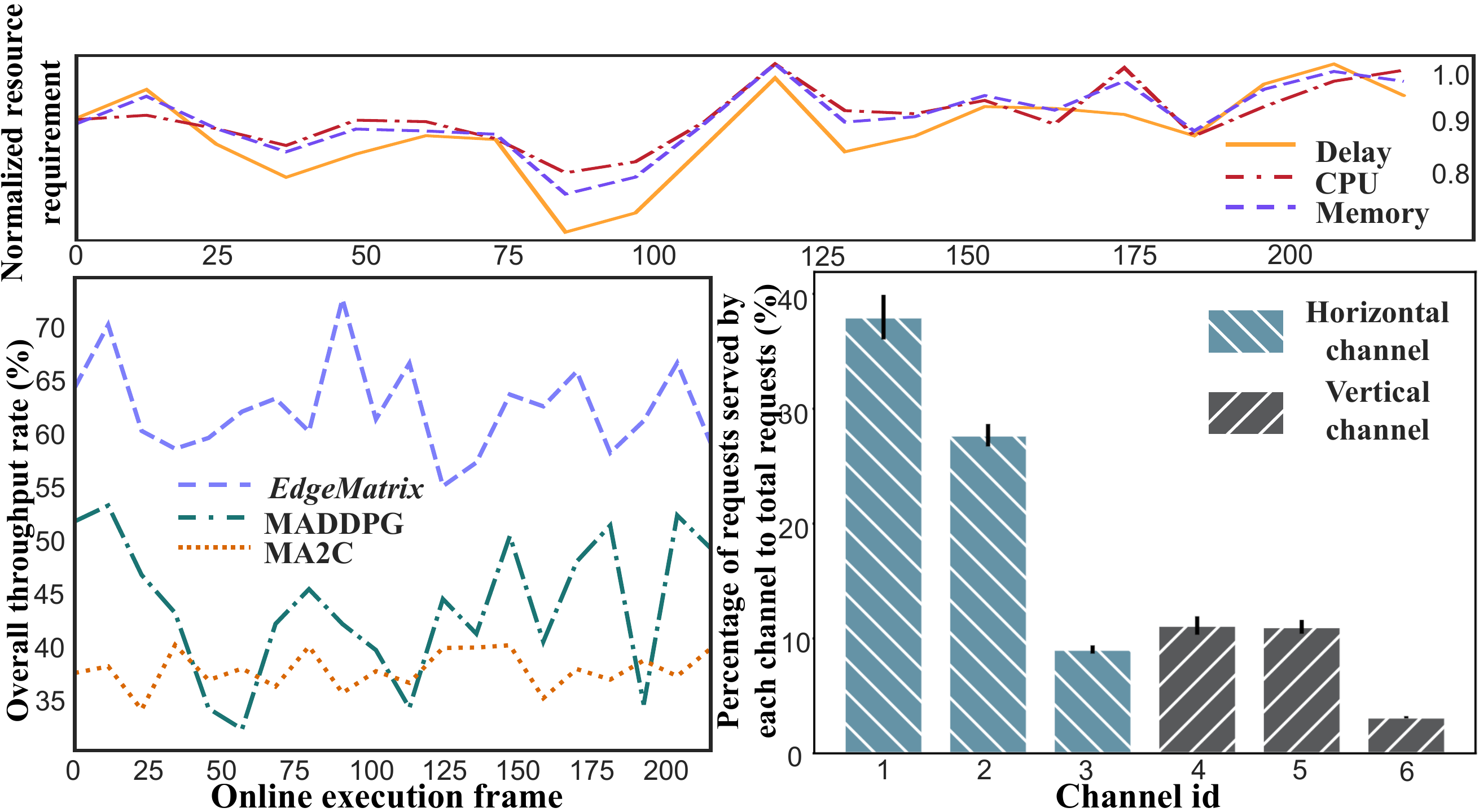}
	\vspace{-1.6em}
	\caption{(a) Under the stochastic request arrivals (top), (b) the overall throughput rate of \textit{EdgeMatrix} against baselines (left buttom), and (c) the percentage of requests served by each channel to total requests (right buttom).}
	\label{fig:Practicability}
	\vspace{-0.6em}
\end{figure}

Fig. \ref{fig:Practicability} proves two sub-{\color{red}objective}s of \textit{EdgeMatrix}: (\textit{$\romannumeral1$}) maximizing overall throughput; (\textit{$\romannumeral2$}) focusing on reducing SLA violations for various user services. We compare the overall throughput rates of different algorithms in Fig. \ref{fig:Practicability}(b) under the same request distribution (as shown in Fig. \ref{fig:Practicability}(a)) within a period and shows the proportion of service requests served by each channel to all service requests in Fig. \ref{fig:Practicability}(c).

Specifically, the performance of \textit{EdgeMatrix} under the same dynamic request distribution is 36.7\% better than the closest baseline. In all six channels (1-6), the smaller the value of Channel\_Id, the higher the SLA priority that the channel can guarantee. Among them, the cells distributed on the channel with {\color{red}channel id} (1-3) are all horizontal, which means that the orchestrated service has a high SLA priority and almost negligible transmission delay, and the channel with {\color{red}channel id} is (4-6) have the vertical cells. Furthermore, under the weight of edge resources proportion set during our training process (i.e., $\mathcal{\varepsilon}=1.5$), the number of requests served by the horizontal channel accounted for 73.7\% of the total number of requests. The number of requests served by the vertical channel is accounted for 26.3\%. Note that the throughput rate of the different services can be adjusted by $\mathcal{\varepsilon}$.

\subsection{Performance in Complex Environment}



To test the adaptability of \textit{EdgeMatrix} in the edge-cloud system, we evaluated the performance of \textit{EdgeMatrix} under the three inherent challenges: multi-resource heterogeneity, resource competition, and networked system dynamics.

\begin{figure}[t]
	\centering
	\includegraphics[width=8.85 cm]{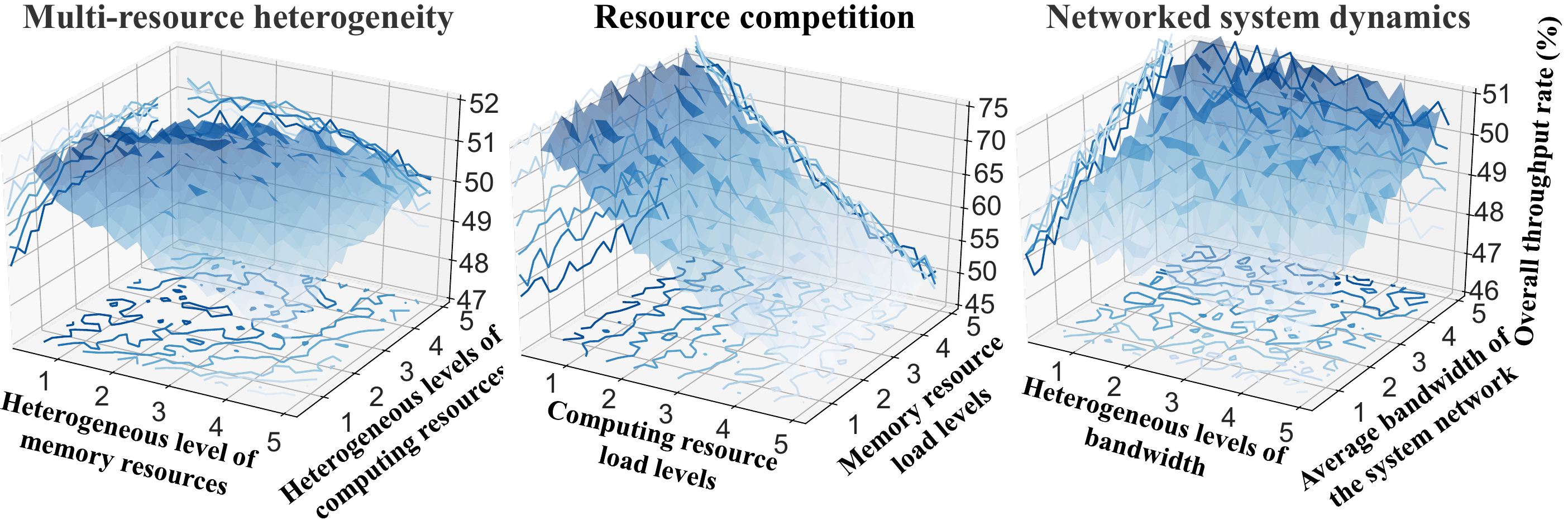}
	\vspace{-1.6em}
	\caption{Practicability of \textit{EdgeMatrix} under (a) hetergeneous resources contains cpu and memory (left), (b) the resource competition (central), and (c) the dynamic networked environment (right).}
	\label{fig:Performance}
	\vspace{-1.6em}
\end{figure}

\textbf{\textit{EdgeMatrix under multi-resource heterogeneity}}.
Fig. \ref{fig:Performance}(a) shows the performance of \textit{EdgeMatrix} under heterogeneous core network (i.e., compute and memory) resources owned by edge nodes. While keeping the total amount of each resource unchanged, we change the resource variance between edge nodes and divide the resource heterogeneity into five levels according to the variance, where the larger the value means the higher heterogeneity. We found that \textit{EdgeMatrix} performs the best when the heterogeneous level of computing resources and memory resources are the same under the existing request distribution. There is a certain correspondence between requests for computing resources and memory resources. Even though the performance of \textit{EdgeMatrix} will decrease as the heterogeneity of one resource increases, the performance of \textit{EdgeMatrix} only drops 3.9\% when the edge node resource heterogeneity is the strongest compared to the weakest case.

\textbf{\textit{EdgeMatrix under resource competition}}.
Resource demand has a significant impact on the throughput rate. The higher the load level of computing resources and memory resources, the more intense the competition for this type of resource. Fig. \ref{fig:Performance}(b) shows that \textit{EdgeMatrix} can adjust adaptively to the dynamic change of resource competition degree, and its ability to adjust the competition of memory resources is better than that of computing resources. \textit{EdgeMatrix} benefits from the isolation ability of channels and the online learning ability of NMAC, which can sensitively perceive the load changes of various resources in the environment and adjust the policy, thus maintain efficient resource customization ability. 

\textbf{\textit{EdgeMatrix under networked system dynamics}}.
The bandwidth resource can affect the stability of the networked system. The heterogeneity level of bandwidth is the same as the previous. The larger the average network bandwidth (1-5), includes larger the average bandwidth. Fig. \ref{fig:Performance}(c) shows that the larger the total bandwidth, the higher the service throughput rate. The system throughput will decrease as the heterogeneity of bandwidth resources between nodes increases. The larger the average bandwidth resource, the smaller the impact of heterogeneity changes. \textit{EdgeMatrix}'s resource customization has played a positive role, which means whether the resources of edge nodes are large or small, \textit{EdgeMatrix} can cover them.

\subsection{Runtime Cost Reduction}

\begin{figure}[t]
	\centering
	\includegraphics[width=7.5 cm]{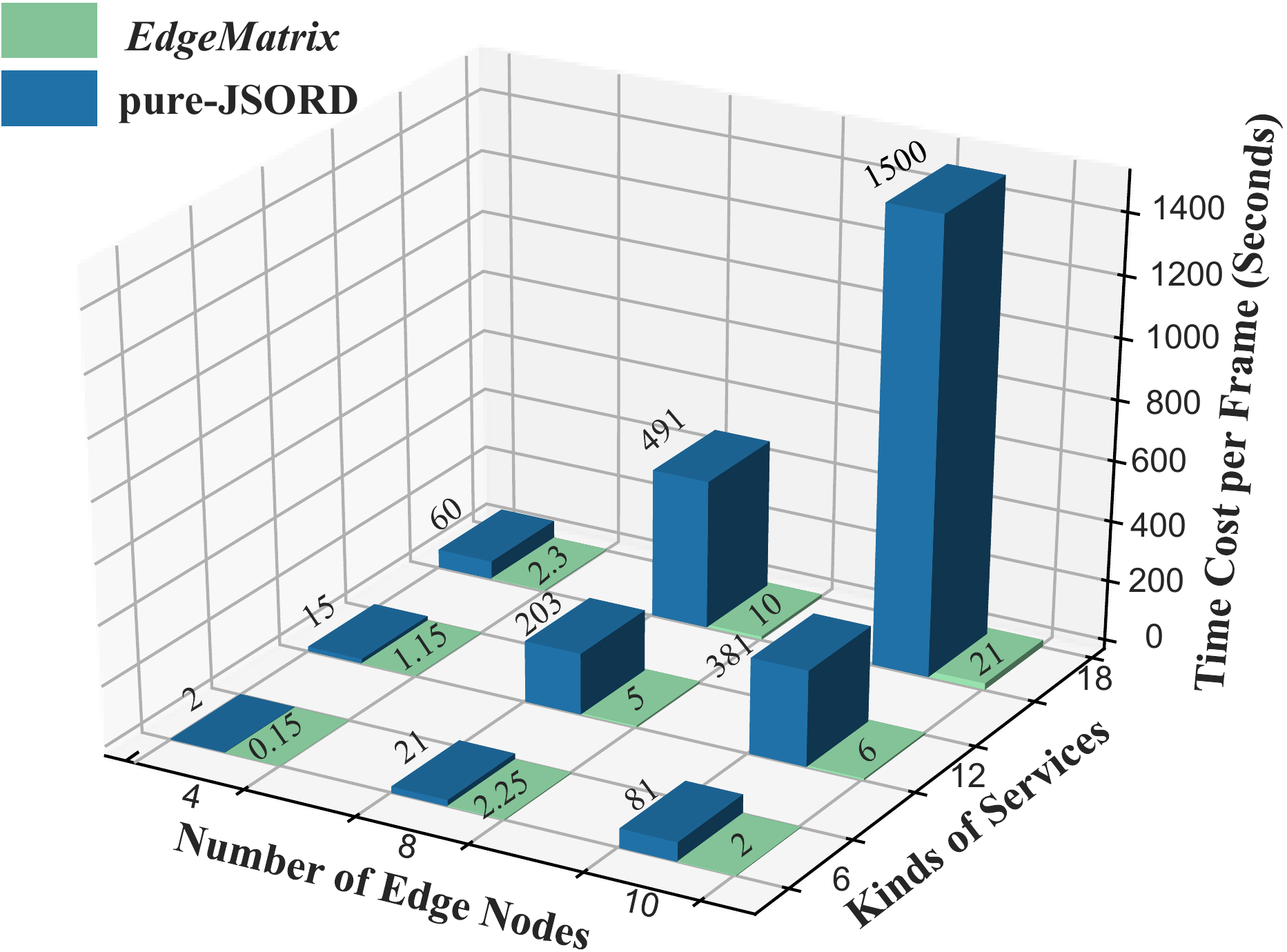}
	
	\vspace{-0.6em}
	\caption{Runtime comparison between \textit{EdgeMatrix} and pure-JSORD.}
	\label{fig:Time Complexity}
	\vspace{-1.2em}
\end{figure}

One of the most important contributions of \textit{EdgeMatrix} is to significantly reduce the {\color{red}runtime} of service orchestration and request dispatch. As shown in Fig. \ref{fig:Time Complexity}, we compare the decision time cost by \textit{EdgeMatrix} and pure-JSORD to perform service orchestration and request dispatch for each frame under different numbers of nodes and service types when the number of channels is 6. We found that the time cost required by \textit{EdgeMatrix} and pure-JSORD will increase with the increase in the number of nodes and service types, but the magnitude of the time cost required by \textit{EdgeMatrix} is much lower than that of pure-JSORD. We observe that the {\color{red}runtime} of pure-JSORD is 13 to 71 times higher than \textit{EdgeMatrix} for a small range of parameter values.
The reason is that the traditional method considers all services and requests within the global nodes, unlike \textit{EdgeMatrix} which (\textit{$\romannumeral1$}) divides the SLA priority levels of user services, orchestrates services with corresponding SLA priority on each channel; (\textit{$\romannumeral2$}) dispatches requests orient to the cells on each channel rather than the global edge nodes, and only dispatch requests with one corresponding SLA priority. These characteristics lead \textit{EdgeMatrix} 
to perform service orchestration and request dispatch in parallel between channels and significantly reduce the magnitude of parameters in the algorithm.


\section{Related Work}
\label{sec:Related Work}



\textbf{\textit{Resource Customization}}. The design concept of network slicing in 5G inspires our work\cite{afolabi2018}, i.e., using SDN and NFV technologies to map resources in physical infrastructure to dedicated virtual resources required by users. Further provide customized services and resource isolation to efficiently utilize limited resources in networked systems, such as RANs\cite{doro2019,mancuso2019slicing,mandelli2019,chen2019,tang2019,hua2019} and Core Network\cite{bega2019,martin2020}. However, some existing research considered a separate MEC host \cite{zhang2019, chantre2020} or Service Chain Functions (SCFs) \cite{mancuso2019slicing,luu2020} in the edge node
as a slice for the edge-cloud system. However, they do not fully consider the multi-resource heterogeneity in edge environments.

\textbf{\textit{Joint Service Orchestration and Request Disapatch}}.We also need to make rational and efficient use of virtual resources in \textit{EdgeMatrix} through service orchestration\cite{liang2019,kayal2019} and request dispatch\cite{dong2019}. In \cite{liang2019}, the authors designed service orchestration algorithms based on the greedy idea to deploy appropriate services in edge clusters. In \cite{kayal2019}, the authors proposed a service orchestration algorithm using a game-theoretic approximation with energy consumption and communication costs as optimization {\color{red}goal}s and enhance the algorithm's robustness by avoiding the need for a centralized auctioneer. However, \cite{liang2019,kayal2019} have no particular concern on (\textit{$\romannumeral1$}) the impact of service orchestration and deletion on request dispatch and (\textit{$\romannumeral2$}) the competition for resources among different services. The authors in \cite{dong2019} designed the request dispatch algorithm aiming to reduce global energy consumption. However, the work in \cite{dong2019} does not fit the edge-cloud system, which is the collaboration between the network edge and the cloud center.




\vspace{-0.6em}
\section{Conclusion}
\label{sec:Conclusion}
\vspace{-0.6em}
In this paper, we {\color{red}propose \textit{EdgeMatrix} that can guarantee the SLA priority for users in the edge-cloud system under three inherent challenges}. \textit{EdgeMatrix} introduces the NMAC to redefine the physical resources in the edge-cloud system as isolated customized resources, which effectively maximize the system throughput and improves the throughput rate of services with high SLA priority under real trace. We also perform JSORD in parallel on each independent channel, significantly reducing {\color{red}runtime} and making \textit{EdgeMatrix} equally applicable to large-scale networked systems.


\end{document}